# Complex network model of the phase transition on the wealth distributions - from Pareto to the society without middle class.

D. Lande[1], A. Snarskii[1], M. Zhenirovskyy[2].

*A model of distribution of the wealth in a society based on the properties of complex networks has been proposed. The wealth is interpreted as a consequence of communication possibilities and proportional to the number of connections possessed by a person (as a vertex of the social network). Numerical simulation of wealth distribution shows a transition from the Pareto law to distribution with a gap demonstrating the absence of the «middle class». Such a transition has been described as a second-order phase transition, the order parameter has been introduced and the value of the critical exponent has been found.*

The distribution of wealth among the people in many countries is described by the Pareto law [1-2], i.e. the number of persons $N(k)$ depends upon their wealth $k$ according to $N(k) \sim k^{-\alpha}$, where $\alpha$ is a constant.

Pareto distributions (scale free distributions) occur in many fields, for instance, in energy distribution of earthquakes (the Gutenberg--Richter law [3]), word frequency distribution in texts (the Zipf law [4]), size of fragments during material breaking [5], etc. The universality of such power distributions is caused by the fact that they are characteristic for the systems, which formation involves (is related to) clusterization, when the union of the «rich» with the «rich» is more preferable as compared with that of the «rich» with the «poor» or the «poor» with the «poor». The effect of priority of such clusterization was named by R. Merton as the Matthew effect [6], according to the biblical quote – «For unto every one that hath shall be given, and he shall have abundance…».

However, there are economies with wealth distribution that do not follow Pareto law. In such cases the function $N(k)$ has a step (a gap) which can be interpreted as the absence of the middle class [7, 8]. Such an economy seems to be much less acceptable than that one that follows Pareto law.

[1] National Technical University of Ukraine "KPI", Kiev, Ukraine.
[2] Bogolyubov Institute for Theoretical Physics of the NAS, Kiev, Ukraine.
Correspondence author is M.Z.- mzhenirovskyy@gmail.com

In present work we propose an approach based on theory of complex networks [9-11]. Introduction of additional rules for wealth allocation triggers Pareto distribution to distribution without middle class. We described such a transition in terms of the theory of phase transitions [12,13] where we revealed the presence of a threshold, introduced the order parameter, and the critical exponent was calculated.

Here we consider the society as the graph and the wealth as a quantity proportional to degree of a vertex (an individual, a firm, etc.), i.e., to the number of edges possessed by the given vertex. The correctness of the given approach can be illustrated, for example, by the activity of businessmen, commercial travelers, or travel agencies, whose success depends often on the number of their partners.

If we consider the dynamics of connections between vertices and, in particular, assume an occurrence of new (connections), then the vertex degree distribution will obviously depend, to a significant extent, on a threshold which needs to be overcome to establish the connections an analog in the economy is a threshold for the entry of a businessman or a company to one or another market). In a standard version (the BA model [9]) of networking with the Pareto distribution [9, 10], there is no such a threshold, i.e., it equals zero.

The BA model [9] is defined as follows: To start with, the network consists of $m_0$ vertices and no edges. One vertex $i$ with $m$ edges is added at every time step. Each edge of $j$ is then attached to an existing vertex with the probability proportional to its degree, i.e. the probability for a vertex $i$ to be attached to $j$ is $P_i = k_i / \sum_j k_j$. In our simulations, $m_0 = 20$, $m = 1$. Here, we assume that the weight of an edge is 1, 2, or 3 with the same probability 1/3. On Fig. 1a degrees of vertices ranked in the decreasing order are shown (curve 1). The dependence matches to the Pareto distribution with the index $\alpha = 0.63$.

Now let's introduce a quantity $r$ that determines the validity of a new attached edge. An edge is attached to vertex $i$ only in case when following relation (1) is true.

$$k_i \geq r \cdot \langle k \rangle \tag{1}$$

Otherwise, the edge is rejected, and we evaluate next vertex. This procedure is i started after 25-th added vertex. At small $r$, the distribution of vertices over degrees (wealth) remains the Pareto distribution.

We found that there exist such a threshold value $r = r_c$, at which the network cease to be a scale-free. The degree distribution of such a network is not more Pareto distribution, the function $N(k)$ manifests a clearly pronounced gap (see Fig. 1b), whose value further denoted as $\eta$. Fig. 2 shows the dependence of $\eta$ upon the parameter $r$. The value of a threshold is approximately $r_c \approx 0.76$. The presence of the gap $\eta$ can be interpreted just as the absence of the middle class. In our example (see Fig. 1b), there are no vertices with numbers from 4 to 65

The transition from a society having middle class (with the Pareto distribution) to society without the middle class can be described n terms of second-order phase transition theory [12, 13]. For this purpose, we have to introduce the order parameter which is zero in the first phase (no gap is present, Pareto distribution is in force) $r < r_c$ and nonzero in the second phase (gap is present) $r > r_c$. Here, we select the gap $\eta$ as the order parameter in the ranked distribution (see Fig. 1b). According to the theory of phase transitions, the order parameter $\eta$ near the threshold ($(r - r_c)/r_c \ll 1$) behaves itself as a power law function of the proximity to a threshold value of the parameter $r_c$,

$$\eta \sim (r - r_c)^t, \quad r > r_c, \qquad (2)$$

where $t$ is the critical exponent, being the principal numerical characteristic of the phase transition. In language of phase transition from the ferromagnetic phase to the paramagnetic one in metals, $\eta$ is the magnetization, $r$ is the temperature, and $r_c$ is the critical Curie temperature.

In Fig. 3, the order parameter $\eta$ as a function of the proximity to threshold $r_c$ is presented in double logarithmic scales. On the basis pf power law approximation of numerical data acc to (2) the critical exponent $t$ was calculated. The critical exponent value can be estimated as $t \approx 1.3$.

It is worth noting note that if we would choose the edge weight equal to 1 at each step at the construction of a network (rather than 1, 2, and 3, as above), we would get also a power dependence of the order parameter on the proximity to a threshold (2) with a different threshold value, but with a close numerical value of the critical exponent $t$.

Analysis of proposed model of wealth distribution demonstrates uneven wealth (vertex degree) distribution with a gap that can be interpreted as the absence of the «middle class» in investigated society. The gap is related to the second-order phase transition.

The model can be used to describe other phenomena in which a separation of the objects by groups "high" and "low can be clearly seen as, for example, in social groups. Each individual who enters into a new social group has to choose those ones who he will be in contact with. When those contacts occur in a random way, their statistics will correspond to the Pareto Law. As soon as a threshold criterion, similar to that one suggested above, appears in a formation of those contacts, Pareto Law shall be broken. Then in such social environment two groups shall stand apart from each other: "sociable" and "unsociable". A criterion of social selection can work like this. With a higher level of our requirements to the social environment there is a more significant gap between those individuals who take an active part in this environment and those ones who have to watch passively what is going on. A number of individual's contacts can be related to his social significance. It's unlikely that an individual with a small number of contacts in a social environment would act in it as an important player.

**Acknowledgements.** Authors are indebted to Igor Bezsudnov for helpful comments of this work.

**Author Information** Correspondence and requests for information should be addressed to M.Z. (mzhenirovskyy@gmail.com).

Fig. 1. Ranked distribution of vertices over their degrees (wealth). The horizontal and vertical axes give, respectively, the vertex number and its degree. *A* – Pareto distribution ($r < r_c$). 1 - parameter $r = 0$, the number of added vertexes is equal 20000, the number of vertices of the network is $20000 + m_0 = 20020$ (no restrictions, all edges are realized). 2 - parameter $r = 0.6 < r_c$, the number of added vertexes is 20000, as in the first case, but the number of vertices of the network is decreased and equals 13532. *B* – distribution with a gap, $r = 0.9 > r_c$. The number of added vertexes is 20000, the number of vertices of the network is 8215, and the order parameter $\eta = 62$.

Fig. 2. Order parameter $\eta$ (a gap between the rich and the poor) versus the parameter $r$. The threshold value $r_c \approx 0.76$. The number of realizations of the network, over which the averaging was performed, was from 100 to 280 for each value of $r$.

Fig. 3. Log-log plot of the dependence of the order parameter $\eta$ on the proximity to the threshold ($r - r_c$). At the selected value of the threshold $r_c = 0.76$, the data are satisfactorily described by the linear dependence $\log \eta \sim t \cdot \log(r - r_c)$, and the critical exponent $t \approx 1.3$.

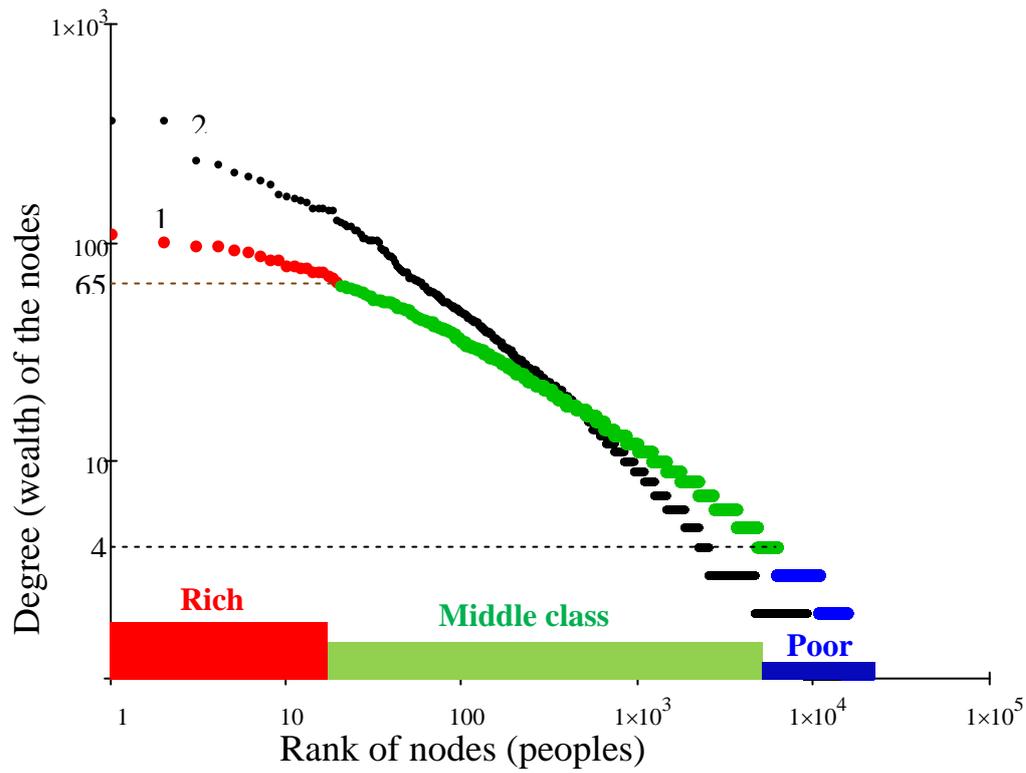

Fig. 1a.

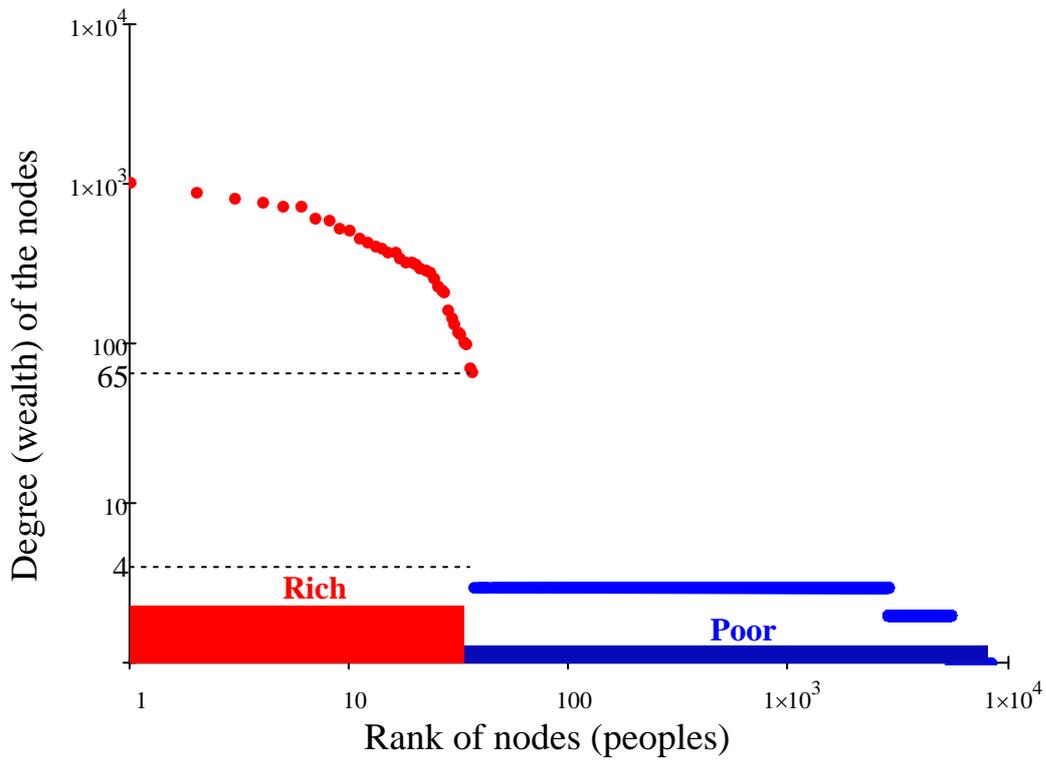

Fig. 1b.

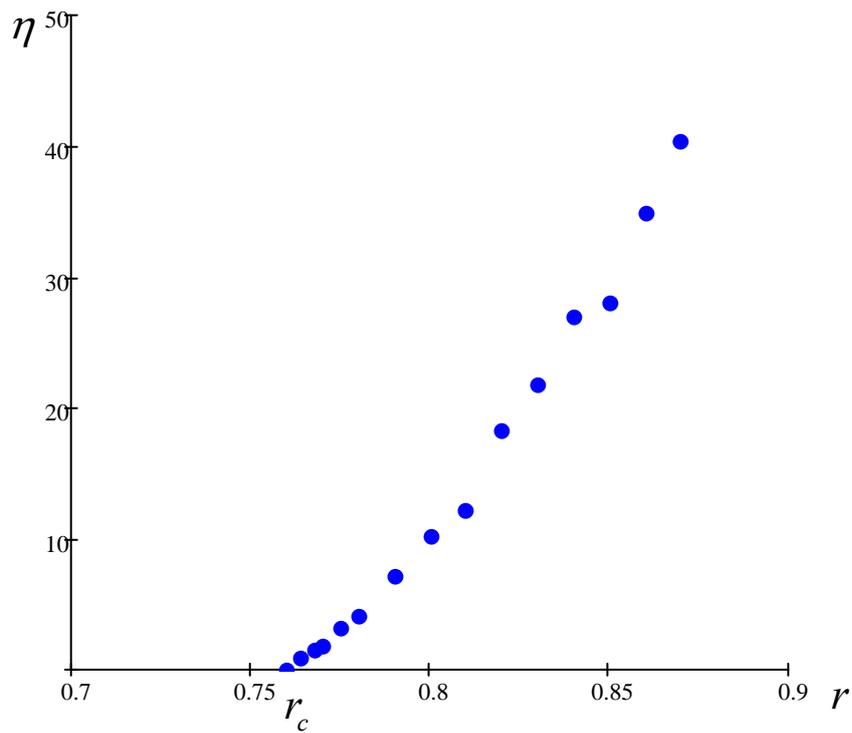

Fig. 2.

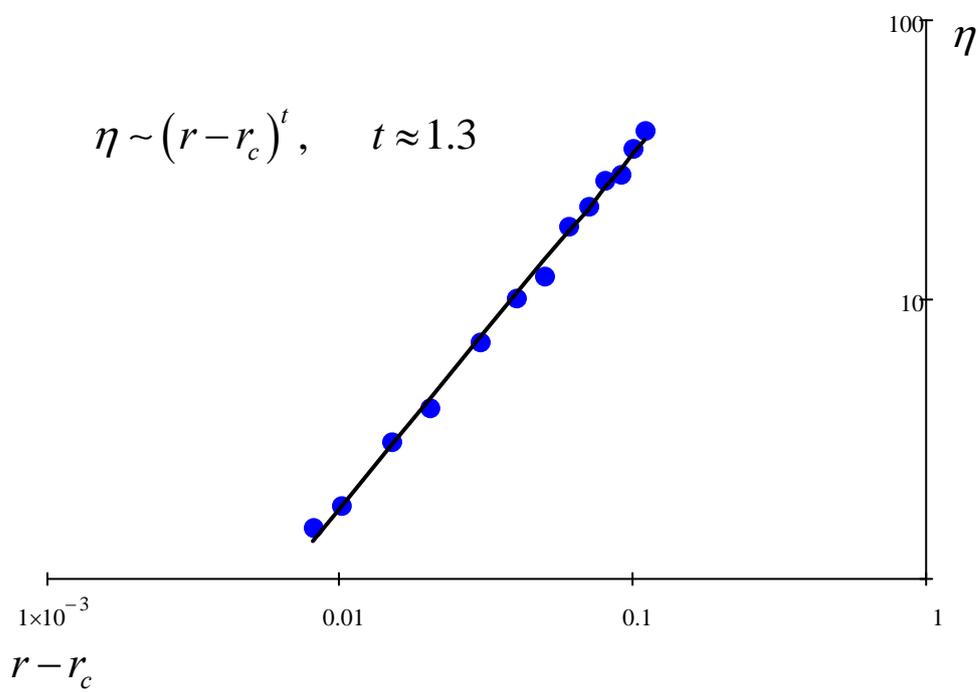

$$\eta \sim (r - r_c)^t, \quad t \approx 1.3$$

Fig. 3.